\documentclass[pra,twocolumn,amsmath,amssymb,showpacs]{revtex4}
\usepackage{graphics}
\usepackage{epsfig}
\newcommand{\bra}[1]{\ensuremath{\left\langle{#1}\right\vert}}

\newcommand{\ket}[1]{\ensuremath{\left|{#1}\right\rangle}}
\newcommand{\ad}{\ensuremath{a^\dagger}}
\def\be{\begin{equation}}       
\def\ee{\end{equation}}
\def\eea{\end{eqnarray}}
\def\bea{\begin{eqnarray}}
\newcommand{\bd}{\ensuremath{b^\dagger}}
\newcommand{\va}[1]{\ensuremath{(\Delta#1)^2}}

\newcommand{\ex}[1]{\ensuremath{\left\langle{#1}\right\rangle}}
\newcommand{\exs}[1]{\ensuremath{\langle{#1}\rangle}}

\begin{document} 
\title{Entanglement detection 
in optical lattices of bosonic atoms with collective measurements}
\date{\today}
\begin{abstract}
The minimum requirements 
for entanglement detection are discussed for a spin chain 
in which the spins cannot be individually accessed.
The methods presented detect entangled states close to a cluster state
and a many-body singlet state, and seem to be viable for experimental 
realization in optical lattices of two-state bosonic atoms.
The entanglement criteria
are based on entanglement witnesses and  
on the uncertainty of collective observables.
\end{abstract}
\author{G\'eza T\'oth}
\affiliation{Theoretical Division, 
Max Planck Institute for Quantum Optics, Hans-Kopfermann-Stra{\ss}e 1, 
Garching, D-85748 Germany.}  
\maketitle

\section{Introduction}

Recently attention has been drawn to optical lattices
\cite{GC03,GC03B,JB99,MG03B}
as promising candidates for the realization of
large scale quantum information processing.
Successful experiments 
have been done by applying state-dependent lattice potentials
for atoms with two internal states.
The lattices were displaced with respect to each other 
and then returned to their original position, making
neighboring sites interact, 
realizing a phase gate and ultimately
a spin chain dynamics \cite{JB99}.
These operations have recently been successfully used to
entangle cold atoms on a large scale \cite{MG03B}. 
As a next step, it is very important 
both theoretically and for quantum
information processing applications to prove 
that the quantum state created is entangled. 
This is, however, a very difficult problem \cite{MG03B}
which has hardly been 
considered (with the exception of Refs. \cite{SM99,B03}).

Entanglement detection in an experiment 
is a hard problem, since
reconstructing the whole density matrix
is usually not possible and the quantum state
is only partially known.
One can typically measure a few observables
yet still would like to detect some of the entangled states.
The situation is even more difficult for 
lattices of two-state atoms created with today's technology
since the lattice sites are not accessible individually \cite{MG03B}.
In this paper we will present scenarios 
where highly entangled states are  detected 
based on very small amount of acquired knowledge: by the measurement 
of collective quantities. Our
schemes are viable with present-day or near-future technology.

The methods to be presented detect entangled states
close to cluster states \cite{RB03,DB03} and many-body singlets.
Cluster states can 
easily be created in a spin chain with nearest-neighbor
interaction \cite{RB03} and have recently been 
realized experimentally in optical lattices \cite{MG03B}. 
They are more immune to decoherence
than other states with genuine multi-qubit entanglement \cite{DB03}
and can be used as a resource 
for measurement based quantum computation \cite{RB03}.
States with total angular momentum zero are also of considerable importance.
One example of such a many-body singlet state is
a chain two-qubit singlets which can serve
as a resource for teleportation and quantum communication.
A singlet of two large spins 
has already been studied in a photonic system \cite{CS03}.
Four-qubit singlets have recently been created with photons
for decoherence free quantum information processing \cite{BE04}.
Optical lattices arise naturally as candidates for realizing
many-body singlets, for example, as a ground state
of Heisenberg chains. 
Note that neither cluster states nor singlets are detected 
by the spin squeezing
criterion \cite{SM99}, which is another approach for 
entanglement detection with global
measurement.

All our results are based on the following simple considerations.
We will build entanglement criteria with 
the three coordinates of the collective angular momentum, $J_{x/y/z}$.
These quantities can be obtained directly 
by population difference measurements,
without the use of multi-qubit operations.
There are now two approaches for entanglement detection:

(i) Entanglement can be detected by measuring 
only $\exs{J_{x/y/z}}$ if the collective measurement is preceded
by some multi-qubit quantum dynamics.

(ii) Without preceding dynamics
an entanglement criterion must involve
second or higher order moments of the
angular momentum coordinates. 
However, an entangled state (e.g., a cluster state)
cannot be detected this way if there
exists a separable state giving the same values for the moments
$\exs{J_{x/y/z}^m}$.

In this paper three necessary conditions 
for separability will be presented.
If these are violated then the system is entangled.
The first one is based on an
entanglement witness, i.e. 
a criterion linear in expectation values \cite{GH02}.  
Connected to it, an experimental scheme
is described to measure the entanglement lifetime 
of a cluster state. This scheme is
viable with present-day technology \cite{MG03B}.
The second method detects also states close to a cluster state,
using uncertainties of collective
observables \cite{OG03,HT02,DG00}. 
In both cases one has to measure
a component of the collective
angular momentum after an evolution under a simple Hamiltonian.
The third method is 
based on measuring uncertainties of all the collective angular momentum
components without a preceding dynamics 
and is a generalization of the approaches of 
Refs. \cite{CS03} and \cite{HT02} for detecting many-body singlets.
In the following we will use the notion of spin chains and lattices of
two-state atoms interchangeably.

\section {Entanglement detection with a witness operator} 

In this section we will show that for all separable states,
i.e. states that can be written as
\begin{equation}
\rho = \sum_l p_l \rho_l^{(1)} \otimes 
\rho_l^{(2)} \otimes ...\otimes \rho_l^{(N)},
\label{sep}
\end{equation}
in a chain of $N$ qubits the following expression
involving third order correlations is bounded from above:
\begin{equation}
J:=\ex{\sum_{k=1}^N \tilde{\sigma}_x^{(k)}} \le \frac{N} {2},
\label{ineq}
\end{equation}
where $\tilde{\sigma}_x^{(k)}=\sigma_z^{(k-1)} \sigma_x^{(k)}
\sigma_z^{(k+1)}$. Here for the end of the chain 
$\sigma_z^{(0)}=\sigma_z^{(N+1)}=1$ and
for simplicity, $N$ is taken to be even.
Later it will be shown how the left hand side of 
Eq. (\ref{ineq}) can be measured as the $x$ component of the collective 
angular momentum after an evolution under a simple
Hamiltonian. 

In order to prove criterion (\ref{ineq}),
first it will be proved that for a product state
\begin{eqnarray}
J_k&:=&\ex{\tilde{\sigma}_x^{(k)}+\tilde{\sigma}_x^{(k+1)}}
\nonumber\\ &=&\ex{\sigma_z^{(k-1)} \sigma_x^{(k)} \sigma_z^{(k+1)}}
+  \ex{\sigma_z^{(k)} \sigma_x^{(k+1)} \sigma_z^{(k+2)}} \nonumber\\
&\le& \left|\ex{\sigma_x^{(k)}}\ex{ \sigma_z^{(k+1)}}\right|+
\left|\ex{\sigma_z^{(k)}}\ex{ \sigma_x^{(k+1)}}\right|\nonumber\\
&\le& 1.
\label{ineq2}
\end{eqnarray}
Note that $J_k$ involves a quadruplets of spins
$(k-1)$, $(k)$, $(k+1)$ and $(k+2)$.
The upper bound  in Eq. (\ref{ineq2}) 
was found using the Cauchy-Schwarz inequality and knowing that 
$\ex{\sigma_x}^2+\ex{\sigma_z}^2\le 1$.
This bound holds for any product state, and since
$J_k$ is linear in expectation values, it also holds for any
separable state.

The upper bound for $J$ is then
$J = \sum_k J_{2k+1} \le N/2$.
(Note that this sum involves $N/2$ overlapping quadruplets 
corresponding to $J_1$, $J_3$, $J_5$, ...)
This proves criterion (\ref{ineq}).
The upper bound in Eq. (\ref{ineq}) is also the lowest possible,
since the separable state $\ket{+1}_x\ket{+1}_z\ket{+1}_x\ket{+1}_z...$
saturates the inequality.

Next a lower bound of the number of entangled qubit quadruplets
will be deduced
from the degree of violation of criterion (\ref{ineq}).
If the quadruplet of spins
$(k-1)$, $(k)$, $(k+1)$ and $(k+2)$ is separable, then
$-1\le J_k\le1$. If it is entangled, then
$-2\le J_k\le2$. Hence a lower limit for
the number of entangled overlapping quadruplets is 
$J-N/2$. The minimum number of 
non-overlapping entangled quadruplets is
half this: $N_{q} \ge J/2 -N/4$.

The criterion (\ref{ineq}) is maximally violated {\it only} for a 
cluster state ($J=N$).
This state is defined as the eigenvector of the following three-qubit
operators
\begin{equation}
\sigma_z^{(k-1)} \sigma_x^{(k)}
\sigma_z^{(k+1)} \ket{\Psi}= \lambda_k \ket{\Psi},
\label{eigen}
\end{equation}  
where $1\le k\le N$ and $\lambda_k \in \{ -1,+1 \}$. We detect a 
cluster state with $\lambda_k=+1$ for all $k$'s. 


The spin squeezing criterion \cite{SM99} does not detect cluster
states as entangled. This criterion is based on the 
necessary condition for separability 
$N\va{J_{\vec{n}_1}}/(\ex{J_{\vec{n}_2}}^2+\ex{J_{\vec{n}_3}}^2)\ge 1$,
where $J_{\vec{n}_k}$ is the total angular momentum
in the direction $\vec{n}_k$ and 
the $\vec{n}_k$'s are perpendicular to each other.
The state is not detected since
for cluster states $\ex{J_{\vec{n}}}=0$ for any $\vec{n}$.

Now we will discuss how to measure $J$. 
It is known \cite{RB03}, that  
$U_{PG}\sigma_x^{(k)} U_{PG}=\sigma_z^{(k-1)} \sigma_x^{(k)}
\sigma_z^{(k+1)}$,
where $U_{PG}=\exp\big\{i\frac{\pi}{4}\sum_k
(1-\sigma_z^{(k)})(1-\sigma_z^{(k+1)})\big\}$ 
denotes an operation implementing a
phase gate for all neighboring spins.
Hence the three-qubit correlation terms
$\tilde{\sigma}_x^{(k)}$ can be measured 
by applying $U_{PG}$ to the chain and then measuring $\sigma_x^{(k)}$.
Also, $J$ can be obtained by applying $U_{PG}$ and then 
measuring the $x$ 
component of the collective spin.
This measurement procedure can only be used to
detect entanglement if the real dynamics of the system 
is known to sufficient accuracy \cite{WM03}.
(However, criterion (\ref{ineq}) can also be used without
a need for multi-qubit dynamics if the particles
are individually accessible. In this case only two measurement settings
are needed for the $\ket{\pm1}_x\ket{\pm1}_z\ket{\pm1}_x\ket{\pm1}_z ...$ 
and $\ket{\pm 1}_z\ket{\pm 1}_x\ket{\pm1}_z\ket{\pm 1}_x ...$ bases.)

Recently, when experimentally creating a cluster state
the effect of decoherence has been observed in the decreasing
visibility of the interference patterns \cite{MG03B}. 
Based on the previous paragraphs, we propose the measurement
of $J$ as defined in Eq. (\ref{ineq}) 
to study the effect of decoherence \cite{DB03}
on many-body entanglement quantitatively.
(Without the many-qubit dynamics
it is hard to observe the decoherence of a cluster state 
since for the whole process $\ex{J_{x/y/z}}=0$.)
The measurement scheme is  
as follows. First a cluster state 
is created starting from $\ket{1111 ...}_x$,
followed by the application of $U_{PG}$.
Then we let decoherence affect the system for time $t_d$. Finally, 
we use $U_{PG}$ again. This would ideally restore the initial state.
However, due to decoherence the measurement of the collective spin 
$J$ in the $x$ direction will give less than the maximal $N$.
The effect of decoherence can easily be followed via the decrease of $J$
with $t_d$.

The influence of a single phase-flip channel acting on spin $(k)$
is given by a completely positive map as
$\epsilon_k \rho = p\rho + 
(1-p)\sigma_z^{(k)} \rho \sigma_z^{(k)}$ where 
$p(t_d)=[1+\exp(-\kappa t_d)]/2$.
Assuming that all these channels act in parallel
one obtains $J(p)/N=2p-1$.
(In the computations
 it was used that $J=N-2$ and $J=N-4$ for 
$\sigma_z^{(k)}\rho_{cl}\sigma_z^{(k)}$
and $\sigma_z^{(k)}\sigma_z^{(l)}\rho_{cl}\sigma_z^{(l)}\sigma_z^{(k)}; k\ne
l$, respectively.) 
The state of the system is detected as entangled by criterion (\ref{ineq}) 
if $p>0.75$. The lower bound for the entanglement lifetime measured
this way is independent of the size of the system,
and as we will show, it is a quite tight bound. 
Following the approach of Ref. \cite{DB03},
one can show that the reduced density matrix of two neighboring qubits 
is entangled if $p>0.71$. 
The entanglement lifetime computed from criterion
(\ref{ineq}) is $20$\% shorter than the lifetime
computed with the latter approach. The difference is slightly
smaller for the partially
depolarizing channel.

\section{Entanglement detection with uncertainty relations}
 
The previous approach detects cluster states with $\lambda_k=+1$
eigenvalues for the defining Eqs. (\ref{eigen}). By modifying
Eq. (\ref{ineq}), one finds that for separable states
\begin{equation}
\sum_{k=1}^N \ex{\tilde{\sigma}_x^{(k)}}^2 \le \frac{N} {2}.
\label{ineq_prime}
\end{equation}
Squaring the expectation value makes it possible to detect
both $\lambda_k=\pm 1$. 
A similar approach for constructing
a nonlinear expression from an entanglement witness 
has been recently presented in Ref. \cite{OG03}.

An expression equivalent to Eq. (\ref{ineq_prime}) 
can be obtained using the variances of
$\tilde{\sigma}_x^{(k)}$: $\sum_k \va{\tilde{\sigma}_x^{(k)}} \ge  N/2$.
Based on this, a collective measurement 
scheme can be defined with the following three
operators
\begin{equation}
X_{1/2/3}:=\sum_{3k+1/2/3} 
\tilde{\sigma}_x^{(k)}.
\end{equation}
$X_{1/2/3}$ is the $x$ component of the collective 
angular momentum operator for every third spin starting from spin $1/2/3$, 
after $U_{PG}$ was executed.
With these, a necessary condition for separability can be obtained
\begin{eqnarray}
\sum_{m=1}^3 \va{X_m} \ge \frac{N} {2}.
\label{sepUnc}
\end{eqnarray}
The proof of Eq. (\ref{sepUnc}) is as follows. 
For a separable state
$\va{X_1}+\va{X_2}+\va{X_3} \ge 
\sum_l p_l \big\{\va{X_1}_l+\va{X_2}_l+\va{X_3}_l\big\}
=\sum_l p_l\sum_k \va{\tilde{\sigma}_x^{(k)}}_l\ge\sum_l p_l \ex{N}_l/2=N/2$.
Here index {\it l} refers to the {\it lth} subensemble.

Note that while the measurement of a 
single operator was enough to
construct the entanglement witness, an entanglement criterion with
observable uncertainties usually involves at least two or three
operators \cite{DG00,HT02,OG03}. 
In our case, distributing the
$\tilde{\sigma}_x^{(k)}$ terms into less then
three operators would make it impossible to use 
$\sum_m \va{X_m}_l=\sum_k \va{\tilde{\sigma}_x^{(k)}}_l$
in the previous derivation.

\section{Entanglement detection when the 
particle number varies on the lattice}

The previous two approaches can straightforwardly be used
for entanglement
detection in optical 
lattices of bosonic atoms with two internal states,
if there is a single atom per lattice site. (However, missing spins
can still be easily handled with these models.)
In practice, it is difficult to prepare a lattice
with unit occupancy \cite{GC03B}.

A method will now be presented which detects
entangled states even if there are several atoms per
lattice site, by measuring collective observables
{\it without} preceding quantum dynamics.
 The necessary condition for separability (proved later) will be
\bea
\va{J_{x}}+\va{J_{y}}+\va{J_{z}}\ge\frac{\ex{N}}{2},
\label{unc}
\eea
where $N$ is the total particle number and
$J_{x/y/z}$ are the collective angular momentum coordinates.
They are the sum
of the corresponding single site Schwinger type angular momentum operators.
For a lattice site, omitting the index $(k)$, 
these are defined as 
$j_x=(\ad b + a\bd)/2$,
$j_y=i(\bd a -a \bd)/2$, and $j_z=(\ad a - \bd b)/2$, where
 $a$ and $b$ are the bosonic destruction operators corresponding to
the two internal states of the atoms.
The particle number at a site is $\ad a+\bd b$.

If the system is in a pure state 
and a lattice site is not entangled with the other sites, 
then its state has the form $\Psi=\sum_{m} c_m \ket{j_m,z_{m}}$.
A separable state is just the convex combination of 
products of such single site states.
Here \ket{j,z} is an eigenstate of 
$j_x^2+j_y^2+j_z^2$ with eigenvalue $j(j+1)$,
and of $j_z$ with eigenvalue $z$.
For example, $\ket{\uparrow}=\ket{1/2,1/2}$ and 
$\ket{\downarrow}=\ket{1/2,-1/2}$
denote a single atom at the lattice site in states $a$ and $b$, respectively,
while $\ket{\emptyset}=\ket{0,0}$ denotes an empty lattice site.

This representation does not take into account entanglement between particles
within the lattice site, as expected, and 
models a lattice site as a particle
with a large spin. 
The spin squeezing criterion \cite{SM99}, however, detects both
entanglement between particles on the same site and entanglement
between particles on different sites.

As we will show, 
criterion (\ref{ineq}) is able to distinguish entanglement due to particle
number variance (e.g., 
$\ket{\uparrow}\ket{\emptyset}+\ket{\emptyset}\ket{\uparrow}$)  
from entanglement in the internal states
(e.g., $\ket{\uparrow}\ket{\downarrow}-\ket{\downarrow}\ket{\uparrow}$ ).
Our aim is to detect the second kind of entanglement.
In the first case 
we have a superposition of states with different on-site particle numbers.
The Schwinger operators commute
with the $N_k$ particle number operators, thus 
by measuring them one cannot
distinguish between a superposition and a mixture of
such states \cite{VC03}. Consequently an entanglement condition in terms
of such observables will not take into account 
entanglement due to particle number variance.

The proof of criterion (\ref{unc}) is based on the relations
\bea
\ex{(j_{x}^{(k)})^2+(j_{y}^{(k)})^2+(j_{z}^{(k)})^2}&=&\nonumber\\
\ex{\frac{N_k}{2}\bigg(1+ \frac{N_k}{2}\bigg)},
\label{statferro}\\
\ex{j_{x}^{(k)}}^2+\ex{j_{x}^{(k)}}^2+
\ex{j_{x}^{(k)}}^2&\le& \frac{\ex{N_k}^2}{4}.
\label{jxyz}
\eea
Here Eq. (\ref{statferro}) expresses the fact, 
that a two-mode bosonic system
has maximal angular momentum \cite{HY00}.
Subtracting Eq. (\ref{jxyz})  from Eq. (\ref{statferro})
 one obtains the uncertainty relation for spin $(k)$ 
\bea
\va{j_{x}^{(k)}}+\va{j_{y}^{(k)}}+\va{j_{z}^{(k)}}
\ge\frac{\va{N_k}}{4}+\frac{\ex{N_k}}{2}.
\label{unc3}
\eea
For separable states
$\va{J_{x}}+\va{J_{y}}+\va{J_{z}}
\ge\sum_l p_l \big\{ \va{J_{x}}_l+\va{J_{y}}_l+\va{J_{z}}_l\big\} 
= \sum_l p_l \sum_k \big\{ \va{j_{x}^{(k)}}_l+
\va{j_{y}^{(k)}}_l+\va{j_{z}^{(k)}}_l\big\}$
which together with Eq. (\ref{unc3}) proves criterion (\ref{unc}).
Thus the uncertainty relations (\ref{unc3}) for the individual lattice 
sites gave a lower bound 
for the uncertainties of the corresponding collective quantities 
for separable states in  Eq. (\ref{unc})\cite{HT02}.
This lower bound is the highest possible, since
{\it any pure product state} with unit lattice site-occupancy 
saturates the inequality. [For atoms on the lattice
a particle number conserving superselection rule applies, 
thus $\va{N_k}=0$ for all pure product states.]

Inequality (\ref{unc}) is maximally violated 
for angular momentum eigenstates with total angular
momentum $J=0$ (many-body spin singlet). 
The spin squeezing criterion \cite{SM99} does not detect these states
as entangled, since they have
$\ex{J_{x/y/z}}=0$.

For two atoms at neighboring lattice sites
such a singlet state is 
$\ket{\Psi_{singlet}}=\ket{\uparrow}
\ket{\downarrow}-\ket{\downarrow}\ket{\uparrow}$.
Chains of two-qubit singlets
of the form
$\Psi_{singlet}\otimes \Psi_{singlet} \otimes \Psi_{singlet}\otimes ...$
also maximally violate our necessary
condition for separability (\ref{unc}). 
In general, many-body singlet states
are ground states of the Hamiltonian $H=J_x^2+J_y^2+J_z^2$.
Maximal violation of inequality (\ref{unc}) can also be obtained 
with the ground state of the antiferromagnetic Heisenberg chain
$H=\sum j_{x}^{(k)} j_{x}^{(k+1)} + j_{y}^{(k)} j_{y}^{(k+1)}
+ j_{z}^{(k)} j_{z}^{(k+1)}$. With a single atom
at each lattice site and for even $N$, 
the non-degenerate ground state
is close to a superposition of chains of two-particle singlets.

It is of experimental interest that
substantial violation of criterion (\ref{unc})
can also be achieved with a  
simple spin chain dynamics with ferromagnetic 
nearest neighbor coupling,
starting out from the state $\ket{\uparrow\uparrow\uparrow ...}$.
Finding the appropriate pulse sequence is a question of numerical
optimization. For example,
$U=\exp(-i\{-3.2\sum j_{x}^{(k)} j_{x}^{(k+1)}-
9.6\sum j_{y}^{(k)} j_{y}^{(k+1)}+
0.8\sum j_{z}^{(k)}\})$ results in a $50\%$ violation
of Eq. (\ref{unc}) for a chain of $N=6$ atoms.

A cluster state is not detected by criterion (\ref{unc}) as entangled.
This is not possible in general with a criterion containing only
the moments $\exs{J_{x/y/z}^m}$. With straightforward 
algebra one can prove that based on the moments for $m\le 4$ 
($m\le 8$), a cluster
state of $N=9$ ($N=17$) particles 
is indistinguishable from the totally mixed state
$\rho_t^{(N)}=\big(\ket{\uparrow}\bra{\uparrow}
+\ket{\downarrow}\bra{\downarrow}\big)^{\otimes N}$. 
The first non-zero moments of $\rho_t^{(N)}$ are $N/4$ and $N(3N-2)/16$
for $m=2$ and $4$, respectively. 

Beside $\exs{J_{x/y/z}^m}$, one might consider the moments of the
more general angular momentum components $J_{\vec{n}}=
\sum_{k=x,y,z} \alpha_k(\vec{n}) J_{k}$. 
However, the following separable state is
indistinguishable from a cluster state based on any such first 
or second order moments: $\rho_s^{(N)}=\exp(i\pi J_y/4)
\big\{\big(\ket{\uparrow\uparrow}_x
\bra{\uparrow\uparrow}+\ket{\downarrow\downarrow}_x
\bra{\downarrow\downarrow}\big)
\otimes\big(\ket{\uparrow\downarrow}\bra{\uparrow\downarrow}+
\ket{\downarrow\uparrow}\bra{\downarrow\uparrow}\big)
\otimes\rho_t^{(N-2)}\big\}\exp(-i\pi J_y/4)$.
Here 
$\ket{\uparrow}_x / \ket{\downarrow}_x = \ket{\uparrow} \pm \ket{\downarrow}$.
Beside having the same moments $\exs{J_{x/y/z}^m}$
for $m\le 2$ as a cluster state, $\rho_s^{(N)}$ also has the same values for 
$A_{kl}=\ex{J_kJ_l+J_lJ_k};
k,l=x,y,z$ ($A_{xy}=A_{yz}=0$, $A_{zx}=1$). 

Thus even if 
there exists an entanglement criterion for cluster states 
based on moments of the collective angular momentum
components, (i) it must involve at least an angular 
momentum component different from $x$, $y$
or $z$, and
(ii) it must be at least third order. This makes the detection of 
cluster states very difficult, if additional many-body dynamics is not
used before the measurement.

\section{Conclusion}

In summary, we have shown how to detect 
entangled states close to a cluster state or a many-body spin singlet 
with collective measurement
in an optical lattice of two-state bosonic atoms. 

\section*{ACKNOWLEDGMENTS}

We would like to thank H.-J. Briegel, J.I. Cirac, T. Cubitt,     
J.J. Garc\'{\i}a-Ripoll, O. G\"uhne and
M.M. Wolf for useful discussions. We also acknowledge the
support of the European Union (Grant No. MEIF-CT-2003-500183), 
the EU projects RESQ and QUPRODIS, and the Kompetenznetzwerk
Quanteninformationsverarbeitung der Bayerischen Staatsregierung.


\end{document}